\documentclass[sigconf]{acmart}

\newcommand{\prototype}{Walking Talking Stick}

\usepackage{graphicx}
\usepackage{caption}
\usepackage{subcaption}
\usepackage{multirow}
\usepackage{dcolumn} 

\newcolumntype{d}[1]{D{.}{.}{#1}}

\AtBeginDocument{%
  \providecommand\BibTeX{{%
    \normalfont B\kern-0.5em{\scshape i\kern-0.25em b}\kern-0.8em\TeX}}}

\copyrightyear{2023}
\acmYear{2023}
\setcopyright{rightsretained}
\acmConference[CHI '23]{Proceedings of the 2023 CHI Conference on
Human Factors in Computing Systems}{April 23--28, 2023}{Hamburg, Germany}
\acmBooktitle{Proceedings of the 2023 CHI Conference on Human Factors in Computing Systems (CHI '23), April 23--28, 2023, Hamburg, Germany}
\acmDOI{10.1145/3544548.3580986}
\acmISBN{978-1-4503-9421-5/23/04}

\begin{document}


\title{The Walking Talking Stick: Understanding Automated Note-Taking in Walking Meetings}

\begin{teaserfigure}
\centering
    \includegraphics[width=\linewidth]{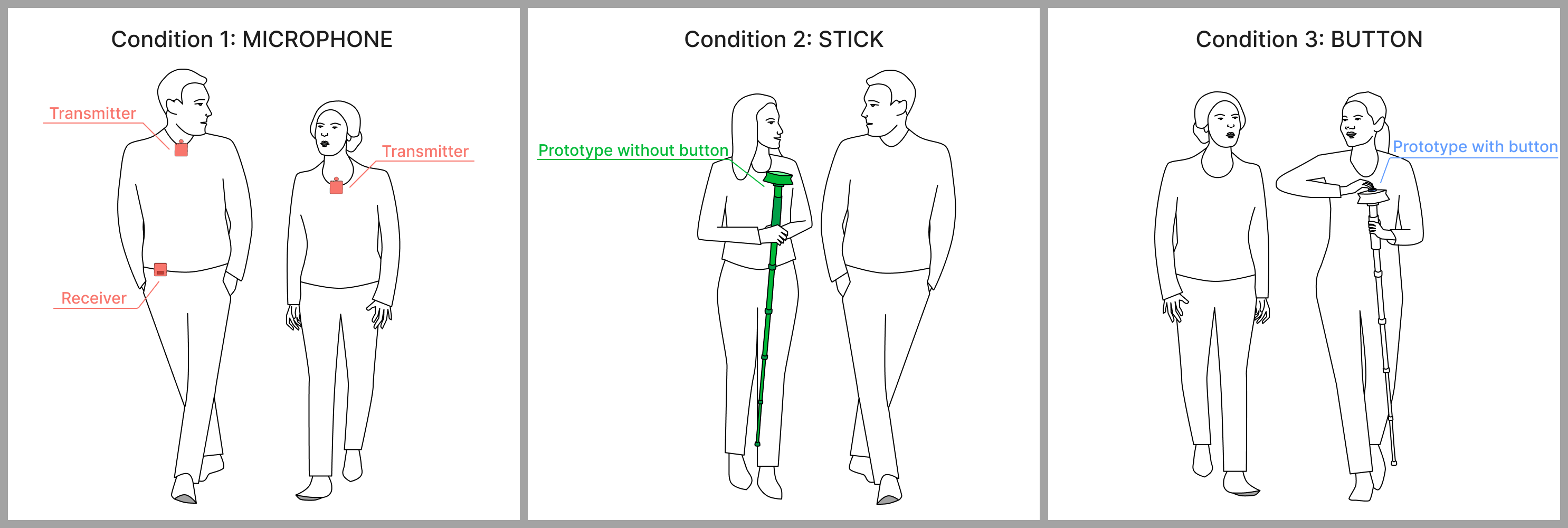}
\Description{}
\caption{We investigated three ways of interacting with automated note-taking while walking. Participants conducted 15-minute walking meetings with clip-on microphones (\textsc{Microphone}), a walking stick (\textsc{Stick}), or a walking stick with a highlighting button (\textsc{Button}). The prototype introduced new conversation dynamics, and the \textsc{Button} facilitated turn-taking and created more useful notes.}
\Description{We investigated three ways of interacting with automated note-taking while walking. Participants conducted 15-minute walking meetings with clip-on microphones (\textsc{Microphone}), a walking stick (\textsc{Stick}), or a walking stick with a highlighting button (\textsc{Button}). The prototype introduced new conversation dynamics, and the \textsc{Button} facilitated turn-taking and created more useful notes.}
\label{fig:teaser}
\end{teaserfigure}

\author{Luke Haliburton}
\email{luke.haliburton@ifi.lmu.de}
\orcid{0000-0002-5654-2453}
\affiliation{%
  \institution{LMU Munich}
  \streetaddress{Frauenlobstr. 7a}
  \city{Munich}
  \country{Germany}
  \postcode{80337}
}

\author{Natalia Bart\l{}omiejczyk}
\affiliation{%
  \institution{Lodz University of Technology}
  \city{Lodz}
  \country{Poland}
}

\author{Pawe\l{} W. Wo\'{z}niak}
\affiliation{%
  \institution{Chalmers University of Technology}
  \city{Gothenburg}
  \country{Sweden}
}

\author{Albrecht Schmidt}
\orcid{0000-0003-3890-1990}
\affiliation{%
  \institution{LMU Munich}
  \city{Munich}
  \country{Germany}
}

\author{Jasmin Niess}
\affiliation{%
  \institution{University of St. Gallen}
  \city{St. Gallen}
  \country{Switzerland}
}

\renewcommand{\shortauthors}{Haliburton, Bart\l{}omiejczyk, Wo\'{z}niak, Schmidt, \& Niess}
\begin{abstract}

While walking meetings offer a healthy alternative to sit-down meetings, they also pose practical challenges. Taking notes is difficult while walking, which limits the potential of walking meetings.
To address this, we designed the \prototype{}---a tangible device with integrated voice recording, transcription, and a physical highlighting button to facilitate note-taking during walking meetings.
We investigated our system in a three-condition between-subjects user study with thirty pairs of participants ($N$=60) who conducted 15-minute outdoor walking meetings. Participants either used clip-on microphones, the prototype without the button, or the prototype with the highlighting button.
We found that the tangible device increased task focus, and the physical highlighting button facilitated turn-taking and resulted in more useful notes.
Our work demonstrates how interactive artifacts can incentivize users to hold meetings in motion and enhance conversation dynamics. We contribute insights for future systems which support conducting work tasks in mobile environments.
\end{abstract}



\maketitle

\section{Introduction}
Many modern workplaces are built around sitting---workers sit at their desks completing tasks and conduct meetings either from their desks online or sitting around a meeting table~\cite{haliburton_technologies_2020}. The negative health consequences of sedentary work have been extensively showcased in public health literature~\cite{clemes_office_2014, creasy_energy_2016, ford_sedentary_2012, gibbs_energy_2017,  zhu_healthy_2020}. One recognized solution to this challenge is holding walking meetings, i.e., discussions conducted during a walk~\cite{kling_opportunities_2016}. Walking is well known to increase worker health and wellbeing~\cite{carr_multicomponent_2013, kelly_walking_2018, wang_long-term_2012}, and enables users to simultaneously conduct a productive task. However, walking meetings also pose a number of challenges~\cite{haliburton_charting_2021, damen_understanding_2020}, particularly difficulties in taking notes. Consequently, there is a need to investigate interactive technologies to support note-taking during walking meetings.

At the moment, walking meetings are only seen as appropriate for early brainstorming or informal meetings where note-taking requirements are minimal~\cite{haliburton_charting_2021, damen_understanding_2020}. Our work is motivated by a desire to make a wider range of meetings feasible to be conducted as walking meetings through technology support. Past research has used (often controversial~\cite{yetim2013critical}) persuasive approaches to convince users to conduct more walking meetings through gamification and route guidance~\cite{ahtinen_brainwolk_2016, ahtinen_lets_2017}. \citet{damen_hub_2020} developed `Hubs' that enable walking meeting participants to periodically record notes and share visuals. However, the Hubs require extensive infrastructure support, do not capture notes in real-time, and limit spontaneity. Audio transcription is one strategy that could support users in generating real-time notes on the move. The possibility of recording meetings already exists with ubiquitous recording-capable smartphones, but uptake remains low. Past work in CSCW has shown that highlighting automatically transcribed notes leads to more useful notes and better recall~\cite{kalnikaite_markup_2012}. Therefore, highlighting (i.e., marking and extracting key phrases in a meeting transcript for faster search and recall) should be explored in the mobile context. Consequently, there is a need to investigate mobile interaction with real-time audio transcription and highlighting to generate useful walking meeting notes.

To address this gap, we present an exploration of automated transcription-based note-taking for walking meetings using a research-through-design approach~\cite{zimmerman_research_2007}. Based on prior research showing that tangible artifacts facilitate memory cues~\cite{bexheti_memstone_2018, tan_kinesthetic_2002} and shared imagination~\cite{baranauskas_tangible_2017},  and that highlighting produces more useful notes~\cite{kalnikaite_markup_2012}, we designed two versions of a prototype called the \prototype{}. 
Both prototypes are recording devices in the form of a tall walking staff, one with a highlighting button and one without. We aim to investigate the impact of transcription-based notes, a tangible artifact, and a highlighting button on conversation dynamics and user experience.
In a between-subjects mixed-method study, thirty pairs of participants ($N=60$) conducted planning meetings while walking outdoors using one of three devices: the \prototype{} (\textsc{Stick}), the \prototype{} with a physical button for highlighting key parts of the meeting (\textsc{Button}), and unobtrusive clip-on microphones as a baseline (\textsc{Microphone}). Specifically, we aim to answer the research question: \textit{How does a shared tangible recording artifact impact walking meetings?}

We found that the tangible \prototype{} promoted conversation turn-taking, increased task focus, and generated new conversation dynamics. The physical artifact created a shared understanding that a serious meeting was taking place, but participants showed a preference for different form factors in certain contexts. We also found that the highlighting button creates more useful notes and facilitates summarization strategies. This paper contributes: (1) the design of the \prototype{}, a tangible device for voice recording and highlighting during walking meetings; (2) a user study evaluating two \prototype{} designs compared to unobtrusive microphones; and (3) insights for future systems that support walking meetings.

\section{Related Work}
In this section, we review past findings that provide context for our research. First, we discuss physical activity at the workplace, followed by an exploration of technologies for meetings on the move. We then review HCI literature on walking meetings and finally report on automated technologies to support meetings.

\subsection{Physical Activity at Work}
Walking has well-documented benefits for physical and mental health. Walking improves cardiovascular capacity~\cite{morris_walking_1997} and reduces blood pressure, weight, and risk of disease~\cite{carr_multicomponent_2013}. Walking also improves depression and anxiety symptoms~\cite{kelly_walking_2018}, happiness and overall mood~\cite{wang_long-term_2012}, and further reduces stress when conducted outdoors~\cite{hunter_urban_2019}. Beyond health benefits, walking also has a positive impact on productive measures such as improved discussions~\cite{balter_walking_2018} and creativity~\cite{oppezzo_give_2014}. Further, prior work has extensively reported on the negative impacts of uninterrupted sedentary behavior~\cite{thorp_prolonged_2012, wilmot_sedentary_2012}.

The benefits of walking are recognized in HCI, and research has aimed to encourage movement in the office. \citet{moradi_neat-lamp_2017}, for example, developed a conceptual framework for workplace movement patterns, which they used to design two prototypes: the NEAT-Lamp and the Talking Tree. The NEAT-Lamp is an unobtrusive ambient display that switches on if the user sits for more than 25 minutes. The Talking Tree, a tree-shaped public display, uses color-changing leaves to visualize movement in a certain area of the office. Their results highlight the social aspect of movement. Our focus on walking meetings is an inherently social movement practice, which is in line with their framework. 

A recent survey by \citet{zhu_healthy_2020} emphasizes that infrastructure in the workplace, such as treadmill desks, leads to an increase in physical activity and a decrease in sedentary time. However, infrastructure adjustments are expensive and often impractical, and many workplaces continue to optimize productivity over physical movement~\cite{haliburton_technologies_2020}. A key limitation of many existing strategies to increase workplace physical activity (e.g.,~\cite{cambo_breaksense_2017, luo_time_2018, kucharski_apeow_2016}) is that they reduce the amount of time spent on productive tasks. By focusing on walking meetings, where motion is integrated into productive tasks, our work aims to increase physical activity without reducing productivity.

\subsection{Meetings on the Move}
Early research in Computer Supported Cooperative Work (CSCW) investigated challenges and opportunities associated with workplace mobility~\cite{bellotti_walking_1996}, highlighting that mobility increases creativity and social interactions but also complexity. \citet{ciolfi_social_2012} emphasized that it is important to be intentional when choosing a collaboration location. Other prior work on mobility by \citet{dahlbom_mobile_1998} characterized mobility in terms of visiting, traveling, or wandering. Based on this framework, when a person joins a meeting in a meeting room or on a virtual platform (e.g., Zoom\footnote{\url{https://zoom.us/}}), they are visiting. If the person answers an email while commuting, they are traveling. Finally, if a person joins the Zoom meeting while walking with their mobile phone, they are wandering. Based on this framework, we aim to move meetings from being mainly a visiting activity toward being a wandering activity.

Prior work in mobile meetings (e.g.,~\cite{wiberg_between_2001}) emphasizes the need to consider spaces outside of traditional meeting rooms for collaboration. \citet{wiberg_roamware_2001} developed RoamWare to support spontaneous mobile meetings via personal digital assistants (PDAs). The aim of RoamWare was to support mobile meetings by taking notes, identifying participants, supporting divided attention, and seamlessly integrating into the mobile setting. Our work aims to extend this support for mobile collaboration by exploring user interactions with automatic note-taking support while walking.

\subsection{Walking Meetings in HCI}
Although walking meetings are promising from a health perspective, HCI research on technology-supported walking meetings has been limited. \citet{ahtinen_walk_2016} developed and refined a mobile application called Brainwolk~\cite{ahtinen_brainwolk_2016, ahtinen_lets_2017}, which uses subtle persuasion through prompts and route guidance to encourage walking meetings. The persuasive approach has generally been criticized in the past by the HCI community based on both practical and ethical concerns~\cite{reitberger_persuasive_2012, spahn_and_2012, smids_voluntariness_2012, yetim2013critical}. As such, rather than taking a persuasive approach, we develop technology to support meetings in motion with the goal of making walking meetings a more attractive option that users willingly choose.

Other research in HCI has investigated supporting walking meetings through infrastructure, such as designated walking meeting paths~\cite{damen_lets_2018} and note-taking `Hubs'~\cite{damen_hub_2020}. The Hubs create a network of stand-up desks where walking meeting participants can periodically stop to take notes or share visuals~\cite{damen_hubs_2021}. \citet{damen_understanding_2020} also contributed barriers and drivers for walking meetings based on walking meetings conducted along a marked path. They identified physical activity, environmental cues, and being outside as key drivers, while lack of notes, size limitations, and distractions were key barriers. \citet{haliburton_charting_2021} further investigated requirements for technology-supported walking meetings and emphasized contradictions between technology support and nature, conversation engagement, privacy, and serendipity. From these prior works, we can see that there is a need for technology to be developed that provides increased support for walking meetings. While the existing positive aspects of walking and talking outside are known to users, the lack of ability to take notes during walking meetings is highlighted consistently in prior work~\cite{haliburton_charting_2021, damen_understanding_2020, ahtinen_lets_2017}. Thus, the inability to effectively take notes emerges as a key hindrance to participating in walking meetings that go beyond brainstorming meetings or open discussions. How best to support users in this task remains an open research question, which we aim to investigate in this work.

\subsection{Automated Meeting Technologies}
Walking meetings are usually associated with brainstorming and informal meetings~\cite{ahtinen_lets_2017, damen_understanding_2020, haliburton_charting_2021}, which limits their potential. Brainstorming is just one of sixteen meeting types defined in the workplace meeting taxonomy by \citet{allen_understanding_2014} and only one of the eight meeting classifications proposed by \citet{ward_applying_1995}. In this section, we identify research aimed at enhancing or automating meeting tasks that can inform walking meeting technologies.

Prior work in HCI has investigated technologies to improve meeting productivity and enhance post-meeting information retention. For example, \citet{bexheti_memstone_2018} developed MemStone to capture and share meeting content to create memory cues. As a post-meeting aid, \citet{shi_meetingvis_2018} created MeetingVis to generate a visual meeting summary based on audio recordings. In our study, we also base our approach on audio recordings. Furthermore, our work takes inspiration from \citet{kalnikaite_markup_2012}, who used a button-based technique to identify hotspots in transcripts to create more useful notes. They found that hotspotting increased conversation contributions and improved recall after meetings. Thus, we incorporate a similar button in our work to enhance the usability of post-meeting transcripts.

Due to the difficulty of taking notes while walking, technologies that automatically generate meeting summaries are particularly relevant. Past work has attempted to create automatic meeting minutes~\cite{zhang_automatic_2012, nedoluzhko_towards_2019} and perform automatic meeting segmentation~\cite{dielmann_automatic_2007}, but both approaches highlight the difficulty of creating usable summaries from such unconstrained input. Researchers have also worked to automate parts of meetings, such as by automatically generating prompt images to help idea generation in brainstorming~\cite{wang_idea_2010} or creating automatic action items from transcripts~\cite{mcgregor_more_2017}.  Several companies have developed products focusing on mobile meetings. Otter.ai\footnote{\url{https://otter.ai/}} transcribes conversations via a mobile app and enables users to add notes and highlights. Spot\footnote{\url{https://www.meetwithspot.com/}} similarly provides transcription, note-taking, and highlighting functions while marketing itself towards walking meeting applications. These companies are examples of a trend in the industry towards remote and mobile work, which motivates our own work. While there is a history of systems that offer automated meeting support, it is yet unknown how users prefer to interact with supporting technology during walking meetings. In our work, we use automatic speech-to-text transcription and highlighting to investigate if state-of-the-art automated meeting assistants can support walking meetings.

\section{The \prototype{}: Design \& Implementation}\label{section:implementation}
To examine our research question of how a shared tangible recording artifact impacts walking meetings, we created two versions of the \prototype{}, one with a highlighting button (\textsc{Button}) and one without (\textsc{Stick}), and evaluated them along with a control condition that used clip-on microphones (\textsc{Microphone}). With these three conditions, we aim to investigate the impact of a tangible artifact and the impact of a highlighting button on walking meetings. In the following, we motivate our design decisions and detail the implementation of the prototypes.

\begin{table*}[t]
    \centering
    \begin{tabular}{c|p{0.7\linewidth}}
        \toprule
        \textbf{Design requirement}       &\textbf{Motivation}\\
        \midrule
        Record 2-4 participants           &Walking meetings are typically effective in groups of 2-4~\cite{haliburton_charting_2021}.\\
        
        Generate transcripts              &Note-taking during walking meetings is difficult~\cite{haliburton_charting_2021, damen_understanding_2020}.\\
        
        Shared tangible artifact          &Tangibles help memory~\cite{bexheti_memstone_2018, tan_kinesthetic_2002}, shared imagination~\cite{l_smith_scaffolding_2021}, and shared storytelling~\cite{baranauskas_tangible_2017}.\\
        
        Function as a walking stick        &A walking stick conveys the idea of motion, walking meetings often take place outdoors, and users are motivated by nature~\cite{haliburton_charting_2021,damen_understanding_2020}.\\
        
        Be easily shared                  &Inspired by talking sticks \cite{wolf2003talking, agyekum2002communicative, smith2011single, thorne1980ceremonial, hicks2001baton, goschke1994development}, shared meeting objects can convey the idea of an important conversation.\\
        
        Enable highlighting               &Highlights make transcripts more useful and help recall~\cite{kalnikaite_markup_2012}.\\
        
        \bottomrule
    \end{tabular}
    \caption{Summary of the design requirements for the \prototype{}}
    \label{tab:designreqs}
\end{table*}

\subsection{Design}
We developed a set of initial design requirements for the prototype based on past HCI research, which extensively outlines technology needs for walking meetings~\cite{haliburton_charting_2021,damen_understanding_2020}. The design requirements are summarized in \autoref{tab:designreqs} and detailed in the following.

Past work (e.g., \cite{haliburton_charting_2021,damen_understanding_2020}, highlights that note-taking is a major challenge for walking meetings. As such, the \prototype{} must be able to record audio while walking outdoors and automatically generate notes. The prototype should capture audio from multiple participants since walking meetings are typically conducted with two to four people.

Tangible artifacts have been used in prior research for recalling~\cite{mugellini_using_2007}, capturing, and sharing personal memories~\cite{bexheti_memstone_2018}.
Prior work has also shown that tangibles can foster shared imagination~\cite{l_smith_scaffolding_2021} and enhance shared storytelling~\cite{baranauskas_tangible_2017}. Therefore, we aimed to develop a shared tangible artifact for walking meetings to foster collaboration, conversation, and shared creativity.

The form of the \prototype{} was chosen as an extension of a traditional hiking staff, which communicates the intention of walking outdoors, and a talking stick, which conveys the idea of shared conversation. Talking sticks signify the right to speak in a group and are present in cultures in multiple places around the world. For example, talking sticks are traditionally used by indigenous peoples of the Northwest Coast of North America~\cite{wolf2003talking}, Akan chiefs in Western Africa~\cite{agyekum2002communicative}, Maori tribes in New Zealand~\cite{smith2011single}, Parliament in the United Kingdom~\cite{thorne1980ceremonial} and Canada~\cite{hicks2001baton}, and also in Kindergartens in many western countries~\cite{goschke1994development}. Talking sticks range in size from small handheld items to large staffs. Although the metaphor is not universal, previous work has shown that the concept is easily grasped and implemented by children in Kingergarten~\cite{goschke1994development}. Thus, we adopted the talking stick as a cross-cultural interaction metaphor suitable for a walking meeting. The design intent is for the talking stick to signify that an important conversation is taking place. The talking stick may be shared but does not necessarily need to be used as a governing item for the right to speak. To operationalize the two metaphors, the prototype should function as a real walking stick, which requires that it be height adjustable, comfortable to hold and walk with and withstand repeated contact with variable outdoor surfaces. The prototype should also be light and not attached to an individual user so it can be easily passed back and forth.

As \citet{kalnikaite_markup_2012} demonstrated that a button can be used to indicate important moments in meetings, which subsequently generates more useful transcripts and improves information recall, we built a second version of the \prototype{} which includes a highlighting button. We constructed two versions of the \prototype{}, one with a highlighting button (\textsc{Button}) and one without (\textsc{Stick}), to discriminate between the impact of carrying a shared tangible artifact and the impact of using a shared highlighting button.

\begin{figure}[t]
\centering
\begin{subfigure}{.47\textwidth}
    \centering
    \includegraphics[width=\textwidth]{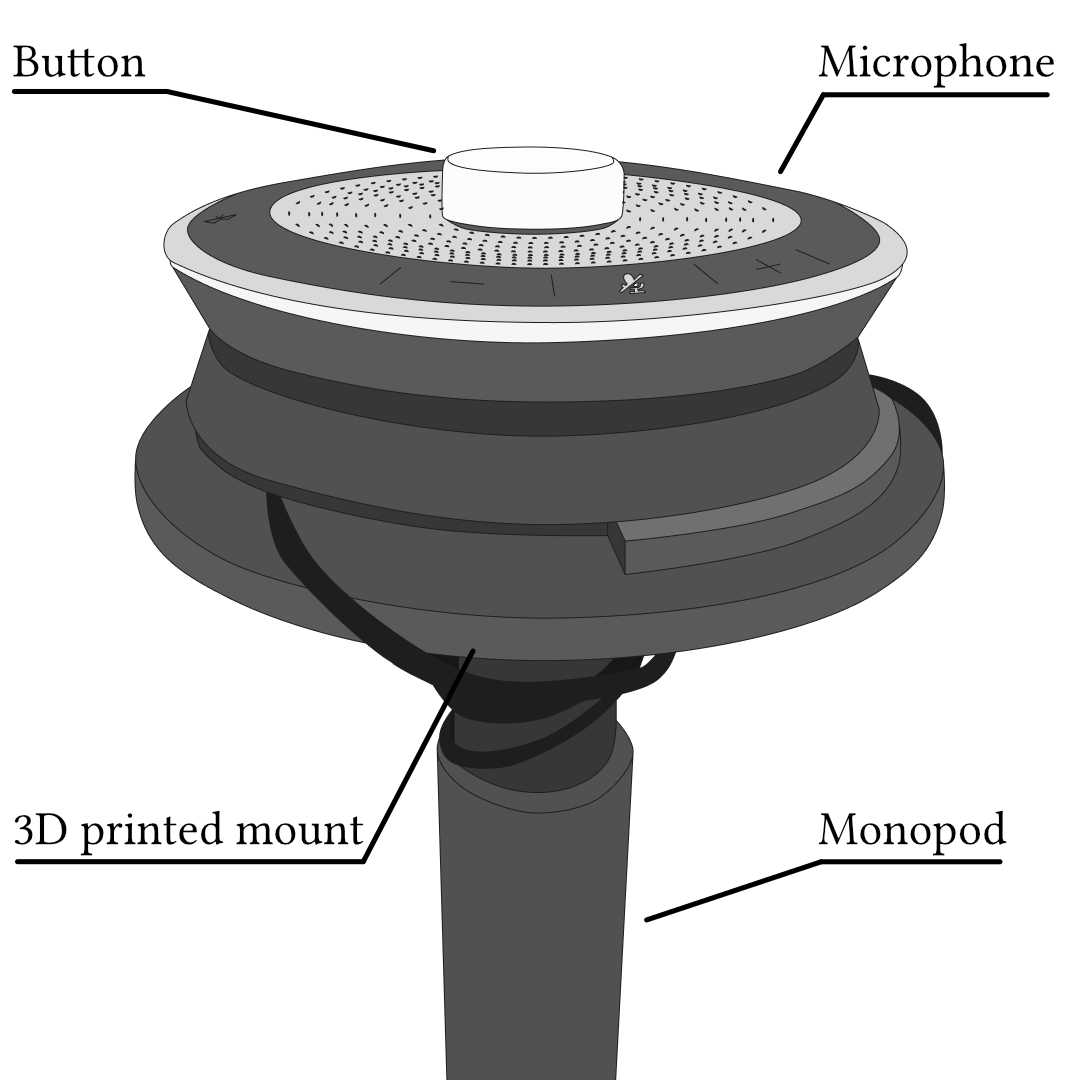}
    \caption{Perspective view of the prototype.}
    \Description{Perspective view of the prototype. A Bluetooth button is mounted on a conference microphone, which is connected to a monopod via a custom 3D-printed component.}
    \label{fig:walkingstickangle}
\end{subfigure}
 \hfill
\begin{subfigure}{.47\textwidth}
  \centering
    \includegraphics[width=\textwidth]{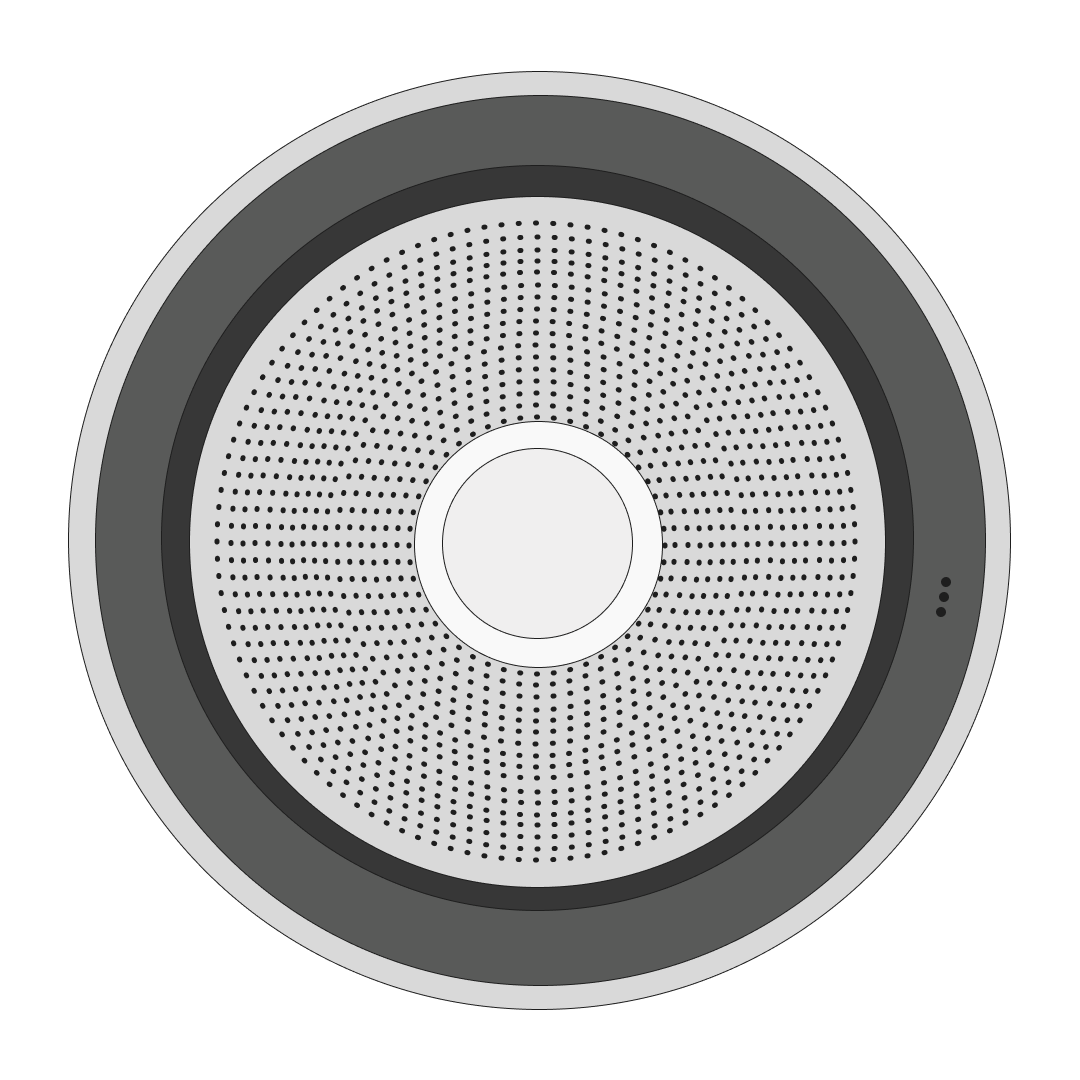}
    \caption{Top view of the prototype.}
    \Description{Top view of the prototype showing the Bluetooth button positioning in the center of the conference microphone.}
    \label{fig:walkingsticktop}
\end{subfigure}
 \hfill
\caption{The \prototype{}. A Bluetooth button is mounted on a conference microphone, which is connected to a monopod via a custom 3D-printed component.}
\label{fig:walkingstick}
\end{figure}

To implement the prototype, we employed an iterative design process. We designed a first prototype using a cane with a 3D-printed extension, a conference microphone mounted to the top and handheld Bluetooth buttons for highlighting. We conducted a pilot study with two pairs of participants. Each pair used the prototype to conduct a 15-minute walking meeting. Based on their feedback, we created a second version of the prototype, shown in \autoref{fig:walkingstick}, with a more comfortable grip, a larger range of height adjustment, a more balanced weight distribution, and an integrated highlighting button.

\subsection{Implementation}

To make it feasible to reproduce the system, we used a combination of off-the-shelf components and 3D printed parts to construct the \prototype{}. The \textsc{Stick} version of the prototype consists of an ENOS SP 30 Bluetooth conference microphone\footnote{\url{https://www.eposaudio.com/en/us/enterprise/products/sp-30-bluetooth-speakerphone-1000223}} mounted to a Sirui AM-306M monopod\footnote{\url{https://www.siruishop.de/en/products/780245}} using a custom 3D printed mounting component\footnote{CAD files are included in the supplementary material}. A standard camera mount thread (1/4'' female) is attached to the center of the 3D printed component using plastic adhesive. The component can then be easily screwed onto the monopod. The microphone slides into the mounting component and is held in place with friction. The conference microphone ensures that we have 360-degree audio capture at a distance of up to 25 m (claimed). We use a monopod because it has a large range of height adjustability, an adjustable rubber- or steel-tipped bottom contact point, a comfortable handle, and balanced weight distribution (it is designed to help photographers to stabilize a camera). We use Otter.ai to record and automatically transcribe conversations in real time on a mobile phone connected to the microphone via Bluetooth.

\begin{figure}[t]
    \centering
    \includegraphics[width=.7\linewidth]{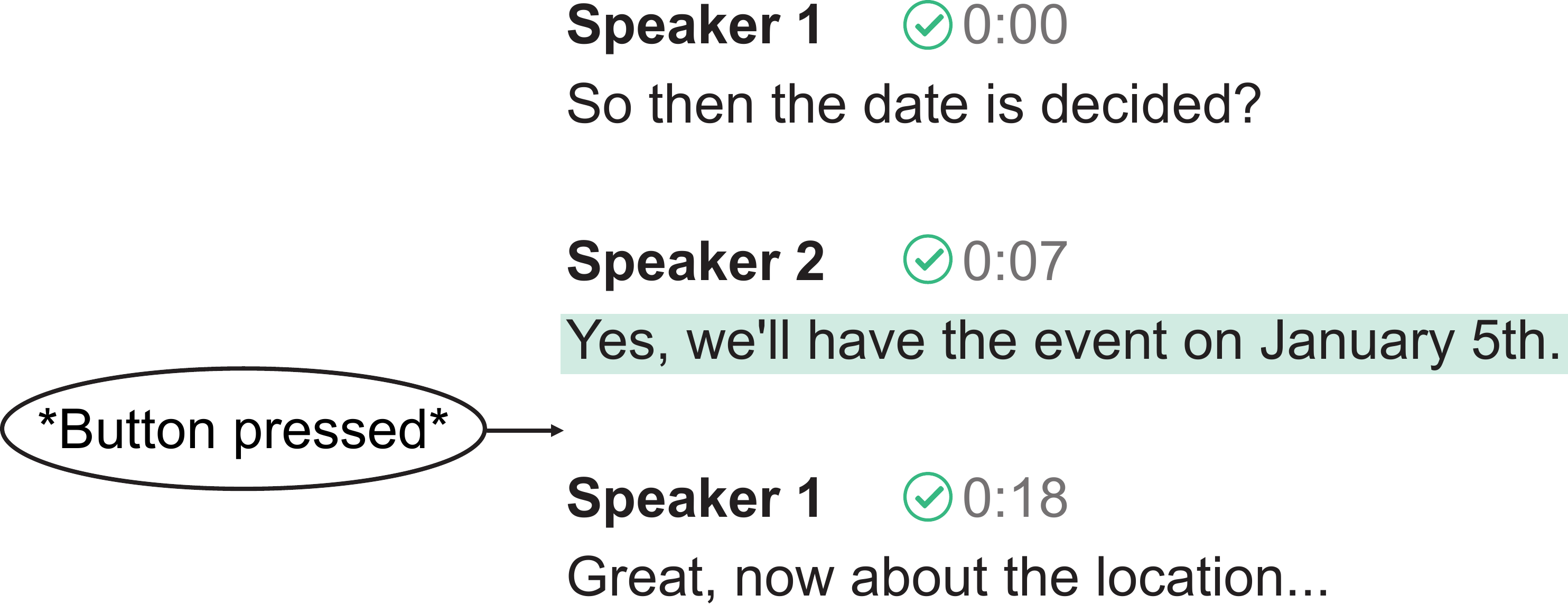}
    \caption{A fictitious conversation demonstrating how the highlighting button functions. One participant presses the button after something important has been said, highlighting the preceding line in the transcript.}
    \label{fig:buttonexample}
\end{figure}

For the \textsc{Button} version of the prototype, we use the full \textsc{Stick} prototype with a Flic Bluetooth smart button\footnote{\url{https://flic.io/}} mounted to the top to facilitate shared highlighting. We use Microsoft Power Automate\footnote{\url{https://powerautomate.microsoft.com/en-us/}} to record timestamps each time the button is pressed. We then use a custom Python script to add highlights at the appropriate timestamps using the highlighting feature built into Otter.ai, which highlights the most recent statement before the button is pressed. As such, users should state something they wish to highlight and then press the button. An example (fictional) conversation is shown in \autoref{fig:buttonexample} to illustrate how the highlighting button functions.

We implemented a third prototype using fully off-the-shelf components as a baseline condition (\textsc{Microphone}). We used two Røde Wireless GO II microphones\footnote{\url{https://rode.com/en/microphones/wireless/wirelessgoii}} connected to a single receiver. As in the \textsc{Stick} and \textsc{Button} prototypes, we use Otter.ai to record and transcribe conversations on a mobile phone connected to the receiver.

These three prototype implementations were developed to investigate two aspects of the prototypes. \textsc{Microphone} is a baseline condition in which the participants are recorded and receive a transcript, but the recording is unobtrusive, and the participants have no shared artifact. The \textsc{Stick} prototype introduces a functional shared artifact and the \textsc{Button} condition adds shared highlighting. In this manner, we can separate the impact of carrying a shared artifact (\textsc{Stick} and \textsc{Button}) and the impact of using a shared highlighting button (\textsc{Button}) from the impact of being recorded (\textsc{Microphone}).

\section{Evaluation}
\begin{figure}[t]
\centering
\begin{subfigure}{.32\textwidth}
    \centering
    \includegraphics[width=\textwidth]{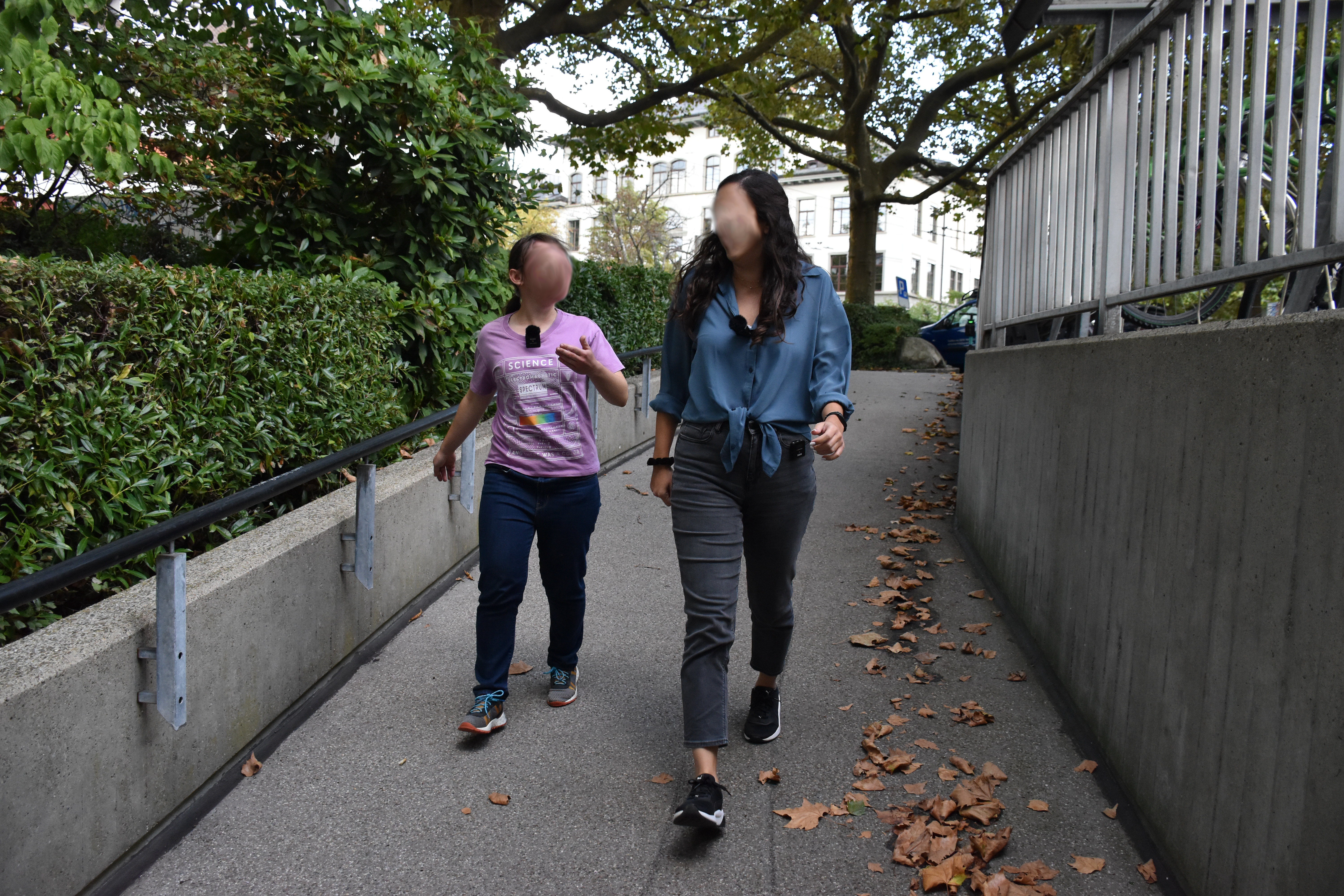}
    \caption{Participants walking with clip-on microphones in the \textsc{Microphone} condition.}
    \Description{Participants walking outside with clip-on microphones in the \textsc{Microphone} condition.}
    \label{fig:conditionm}
\end{subfigure}
 \hfill
\begin{subfigure}{.32\textwidth}
  \centering
    \includegraphics[width=\textwidth]{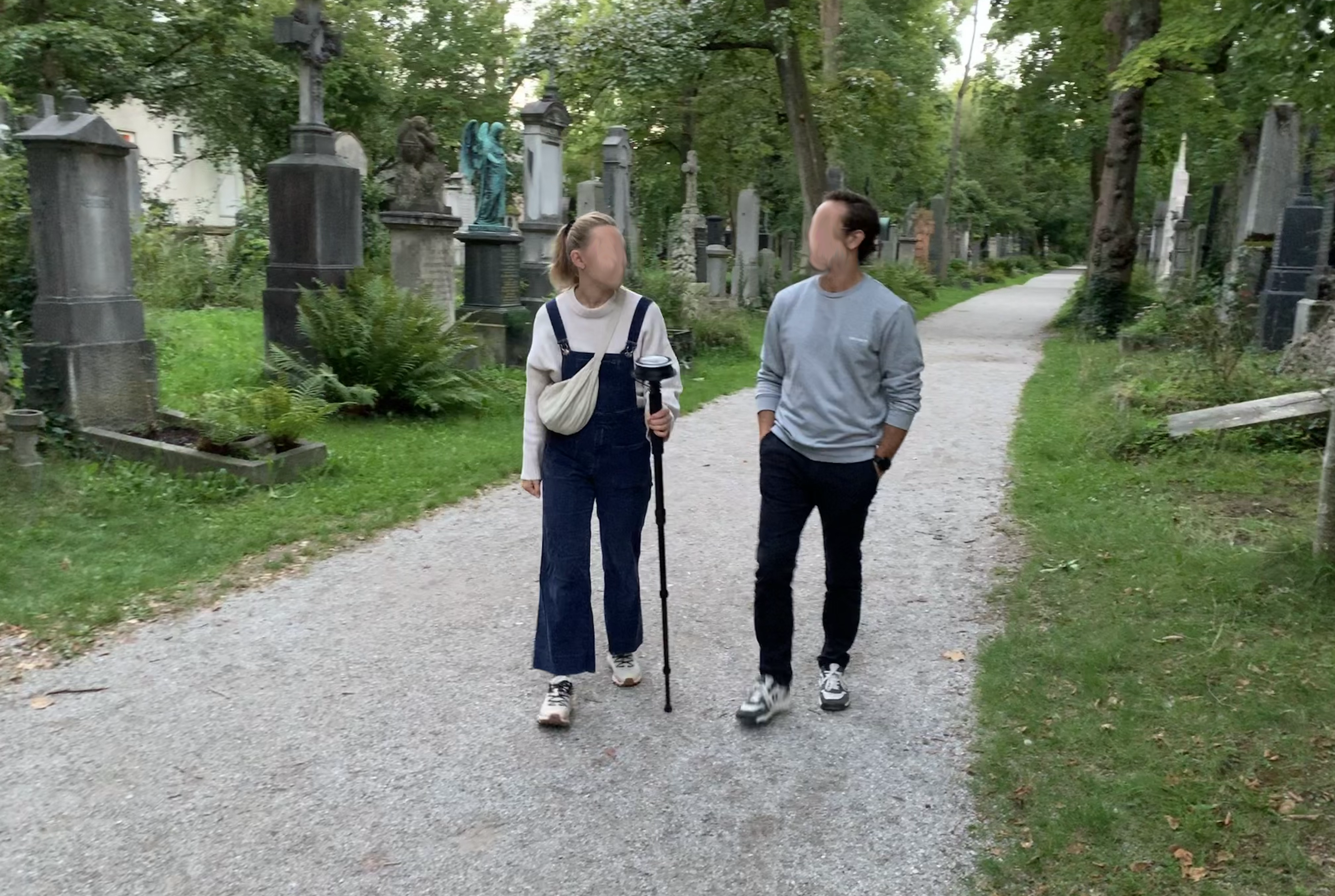}
    \caption{Participants walking with the prototype in the \textsc{Stick} condition.}
    \Description{Participants walking with the prototype in the \textsc{Stick} condition.}
    \label{fig:conditions}
\end{subfigure}
 \hfill
\begin{subfigure}{.32\textwidth}
  \centering
    \includegraphics[width=\textwidth]{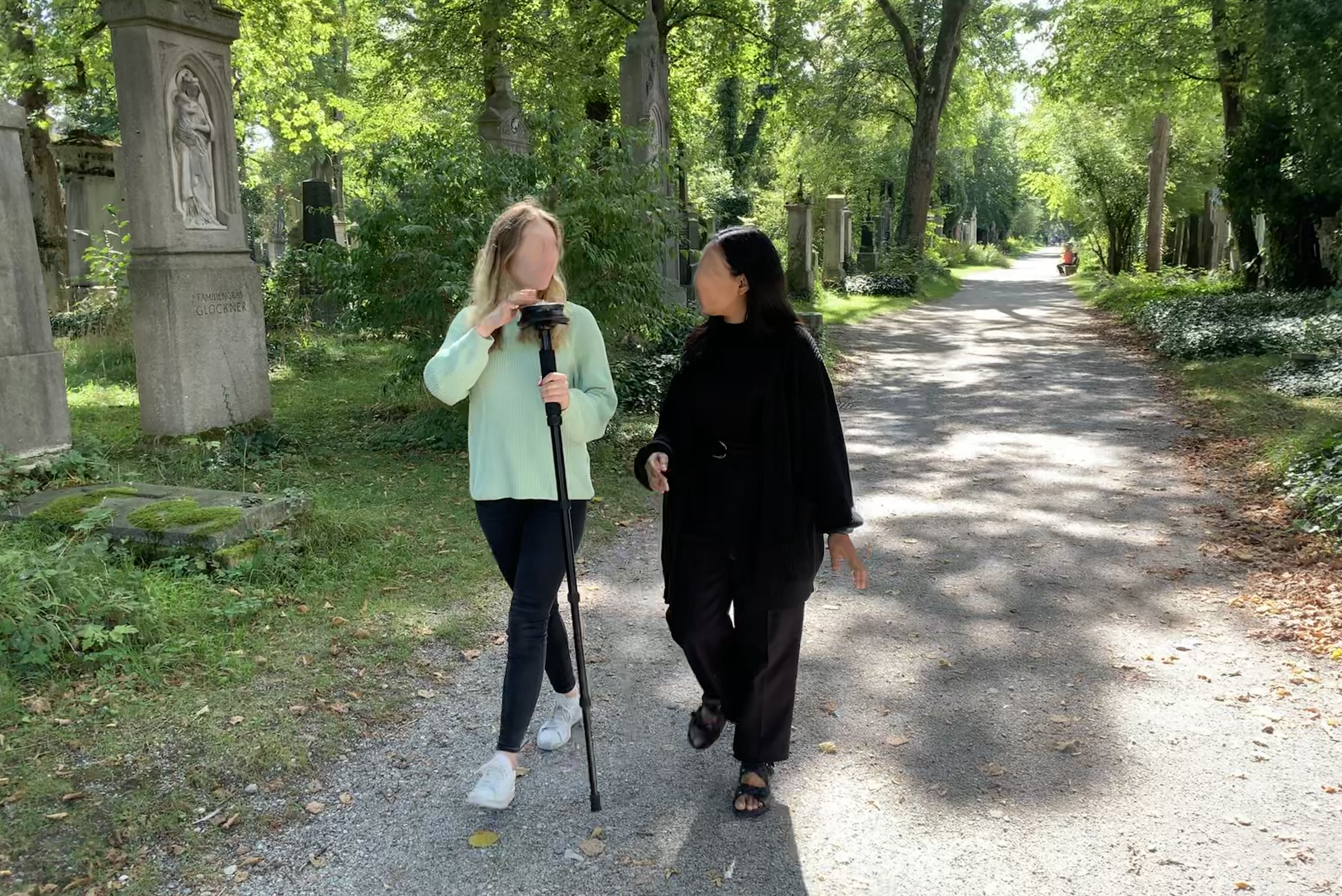}
    \caption{Participants walking with the \prototype{} in the \textsc{Button} condition.}
    \Description{Participants walking with the \prototype{} in the \textsc{Button} condition.}
    \label{fig:conditionb}
\end{subfigure}
\caption{Participants conducting walking meetings in each condition.}
\label{fig:photos}
\end{figure}

We conducted an exploratory between-subjects outdoor user study to investigate automatic note-taking during walking meetings. Participants completed the study in pairs and were randomly assigned to one of three experimental conditions (\textsc{Microphone}, \textsc{Stick}, or \textsc{Button}, described in the following section). Photos of participants in each condition are shown in \autoref{fig:photos}.

\subsection{Participants}
We recruited thirty pairs of participants for a total of $N$=60, aged 20-34, M = 26.55, SD = 3.4, 29 participants identified as female and 31 as male. Participants were recruited using a university mailing list and snowball sampling. We recruited pairs of participants, so each pair already knew one another to some extent. Most (17) pairs of participants knew one another for longer than one year, with seven pairs knowing each other for 6-12 months, three pairs for 1-6 months, and three pairs for less than one month. We compensated participants at a rate of 10€/hour and the experiment took approximately one hour, resulting in a total of 10€ per participant. The study was approved by the ethics committee within the University Faculty\footnote{Details removed for anonymization purposes}. We will refer to participants using their condition (M, S, or B) and study ID (e.g., B05).

\subsection{Task and Conditions}\label{sec:task}
Each pair of participants completed the same task, regardless of their experimental condition. The task was to plan a (fictitious) future workshop during a 15-minute walking meeting. Participants were instructed to conduct the meeting as though they would be required to follow through on their workshop plans. We provided the participants with a one-page Meeting Prompt\footnote{The Meeting Prompt is included in the supplementary material} describing the task. The prompt also contains a list of ``things to consider'', instructing them to choose a meeting topic that combines their areas of expertise, select potential dates, identify speakers and attendees to invite, draft a schedule, and distribute tasks among themselves. These considerations were added to ensure that participants discussed concrete decisions, shifting the meeting towards being a \textit{Planning Meeting} rather than a \textit{Brainstorming Meeting}.

We examined three walking meeting technology conditions, \textsc{Microphone}, \textsc{Stick}, and \textsc{Button}, in a between-subjects experimental design to evaluate our research question:

\subsubsection*{\textbf{\textsc{Microphone}:}} Each participant was equipped with a clip-on microphone. One participant also carried the wireless receiver connected to a smartphone. Clip-on microphones were used because the microphone on a smartphone is not sufficiently powerful to capture audio from two participants from a pocket while walking outside. In this condition, the meeting was recorded by wireless microphones and transcribed by Otter.ai. The recording was active during the entire meeting, and participants reviewed the transcript at the end.

\subsubsection*{\textbf{\textsc{Stick}:}} Each pair of participants was given a single \prototype{} to carry in their meeting. The microphone on the prototype recorded the meeting, and Otter.ai generated transcriptions. The participants were free to choose who carried the \prototype{} throughout the meeting and were allowed to switch carriers. The recording was active during the entire meeting, and participants reviewed the transcript at the end.

\subsubsection*{\textbf{\textsc{Button}:}} Each pair of participants was given a single \prototype{} with a button on top. The microphone on the prototype recorded the meeting, and Otter.ai generated transcriptions. Participants could use the button to generate highlights at any time. The participants were free to choose who carried the \prototype{} and who pressed the button throughout the meeting and were allowed to swap and interchange these roles as they chose. We explained to participants that the highlighting button should be pressed after an important statement is said. The recording was active during the entire meeting, and participants reviewed the transcript at the end.

These three conditions were designed to shed light on several aspects of technology-supported walking meetings. The \textsc{Microphone} condition provides a baseline where participants experience automated note-taking while conducting a walking meeting with solely off-the-shelf components. The \textsc{Stick} condition introduces the prototype to investigate how a shared tangible artifact impacts the user experience and conversation dynamics. Finally, the \textsc{Button} condition investigates shared highlighting because prior work suggests that this generates more useful transcripts~\cite{kalnikaite_markup_2012}. Whether or not there should be highlights was a key design consideration since this requires (potentially distracting) manual interaction with the device instead of only using passive recording, which is why the \textsc{Stick} and \textsc{Button} conditions were separated. We did not include a condition with unobtrusive microphones and highlighting as this would have required designing a second device to input highlights. Adding a button to this condition would introduce a tangible element, and therefore would not be sufficiently different from the \textsc{Button} condition to isolate the button. In all, these conditions constitute a research-through-design approach, which is appropriate to investigate under-constrained problems~\cite{zimmerman_research_2007}.

\subsection{Measures and Analysis}\label{sec:measures}
We aimed to investigate how the \prototype{} impacts conversations and to assess the usability and user experience of the device. As such, we collected data from recorded meeting conversations, questionnaires, and exit interviews. Each of these data sources requires a separate method of analysis. We have included the interview protocol and full questionnaire in the supplementary material.

We analyzed the meeting recordings to understand how the prototype impacts conversations. We based our conversation analysis on past work on turn-taking and conversational dialogues~\cite{roter_interactive_2008, chan_designing_2008}. The recorded meeting conversations were timestamped and labeled with speaker identification by the Otter.ai software. One author listened through each recording and ensured that the labeling was clean. The transcripts for each recording were then analyzed for relevant conversation metrics. In particular, we calculated Turn Density (number of words per speaker turn), Interactivity (number of speaker turns per minute), and Speaking Ratio (speaking time of the dominant speaker divided by the other). A speaker turn is a segment of uninterrupted speech by one speaker~\cite{roter_interactive_2008}. To analyze the conversation metrics, we use ANOVA methods similar to \citet{chan_designing_2008}. Depending on normality, based on Shapiro-Wilk testing~\cite{wobbrock2011art}, we report one-way ANOVA or ART-ANOVA results. Where appropriate, we then use Tukey post hoc tests to report comparisons between groups.

In the questionnaire, we collected responses on usability and user experience. We recorded Likert-scale responses to four questions on user perceptions (stress level during the meeting, how engaged they were in the conversation, how much they felt they contributed to the conversation, and likelihood to use the system in the future). We also recorded responses to two standardized scales. We analyzed the AttrakDiff~\cite{hassenzahl_attrakdiff_2003} and System Usability Scale (SUS)~\cite{brooke_sus_1995} according to their original documentation. We analyzed responses to Likert-type scales using ANOVA procedures (ART-ANOVA) on aligned-rank transformed data as the method is suited to analyzing ordinal data~\cite{wobbrock2011art}. Mean cell frequencies were checked to ensure applicability of the test~\cite{luepsen}. Where applicable, post-hoc pairwise comparisons were performed using the Tukey method.

Finally, to gain a deeper insight into the user's experiences, we also collected qualitative responses through exit interviews. We asked participants about their experience, the dynamics of their conversation (e.g. if anyone took the lead or spoke more), how the presence of the system impacted their conversation, and their strategy for interacting with the device (e.g. who carried the stick, if they swapped, who pressed the button, when they pressed the button). We recorded and transcribed all interviews verbatim. We imported the interview transcripts into Atlas.ti\footnote{\url{https://atlasti.com/}} analysis software. As a first step, four researchers used open coding to code a representative sample of 15\% of the material. Following this, the researchers discussed and agreed on a coding tree. Finally, one researcher coded the remaining material. This process is in line with \citet{blandford_qualitative_2016}.

\subsection{Procedure}
\begin{figure*}[t]
    \centering
    \includegraphics[width=0.8\linewidth]{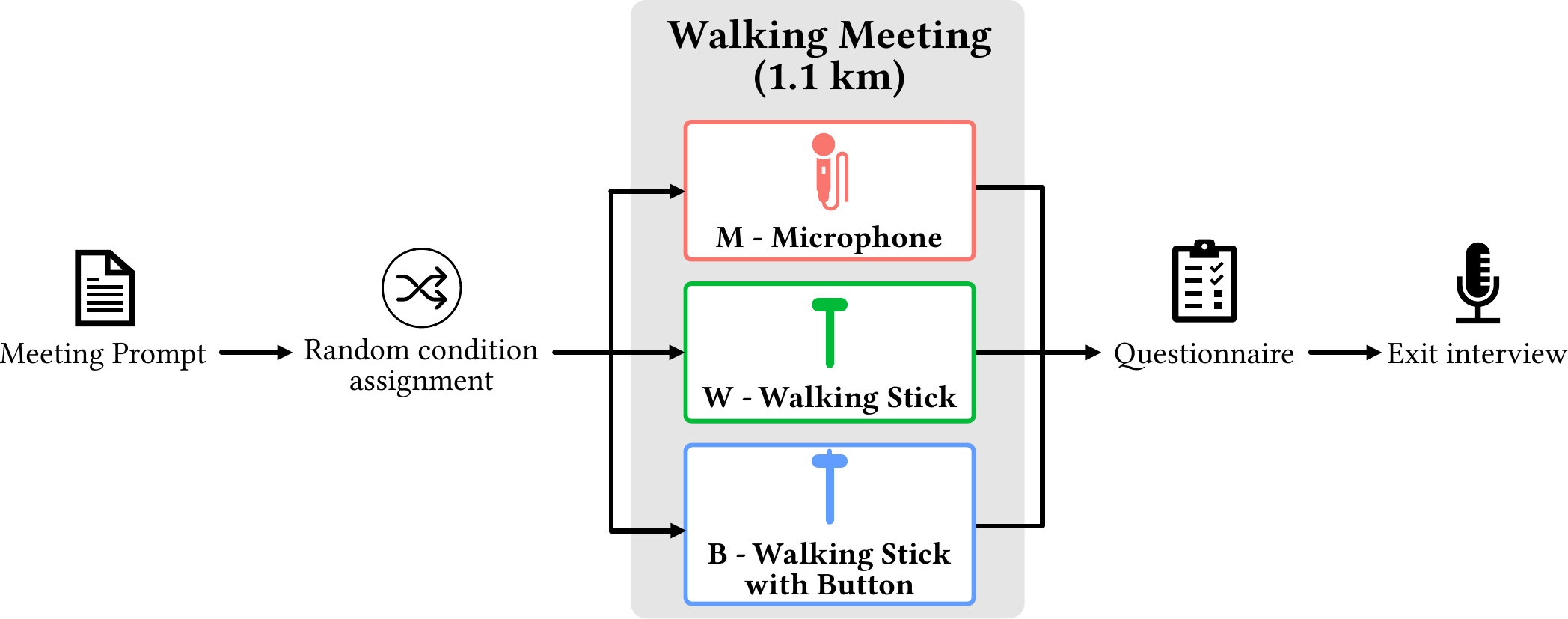}
    \caption{Study Procedure: Pairs of participants are given a one-page Meeting Prompt and are then randomly assigned to either the \textsc{Microphone}, \textsc{Stick}, or \textsc{Button} condition. Each pair then completes a walking meeting on a 1.1 km route based on the Meeting Prompt. Following the meeting, each participant completes a questionnaire and an exit interview.}
    \Description{Study Procedure: Pairs of participants are given a one-page Meeting Prompt and are then randomly assigned to either the \textsc{Microphone}, \textsc{Stick}, or \textsc{Button} condition. Each pair then completes a walking meeting on a 1.1 km route based on the Meeting Prompt. Following the meeting, each participant completes a questionnaire and an exit interview.}
    \label{fig:timeline}
\end{figure*}

A timeline of the study procedure is depicted in \autoref{fig:timeline}. After obtaining informed consent, each participant was provided with a one-page Meeting Prompt and given time to read through and ask questions.

Depending on their study condition, the two participants were then introduced to either the clip-on microphone, the base \prototype{}, or the \prototype{} with the highlighting button. The experimenter then explained the route to the participants, which was a 1.1 km loop in a local park. This distance was selected as it takes approximately 15 minutes to complete. The experimenter initiated the Otter.ai recording and transcription software, and the participants walked around the designated route while conducting a meeting based on the prompt. During the study, the participants were visible for parts of their route. The experimenter observed them when they were visible, allowing them to ask follow-up questions if relevant behaviors were observed (e.g., handing the stick to the other person after a certain time). This was complemented by the interview protocol elaborated in Section \ref{sec:measures} and included in the supplementary material.

After completing the meeting task, participants were given the phone with the transcript to freely review within the Otter.ai app. Participants in the \textsc{Button} condition could also see the highlights they created. The highlighted segments were presented in situ within the transcript (as shown in \autoref{fig:buttonexample}) and as a separate list accessed via a menu button on the side. Each participant participated in a post-study questionnaire and an exit interview. One participant completed the questionnaire while the experimenter engaged the other in an interview, after which the participants switched.

\section{Results}
In the following, we present the results for conversation metrics, user perceptions, usability, and exit interviews.

\subsection{Conversation Metrics}
\begin{figure}[t]
\centering
\begin{subfigure}{.47\textwidth}
    \centering
    \includegraphics[width=\textwidth]{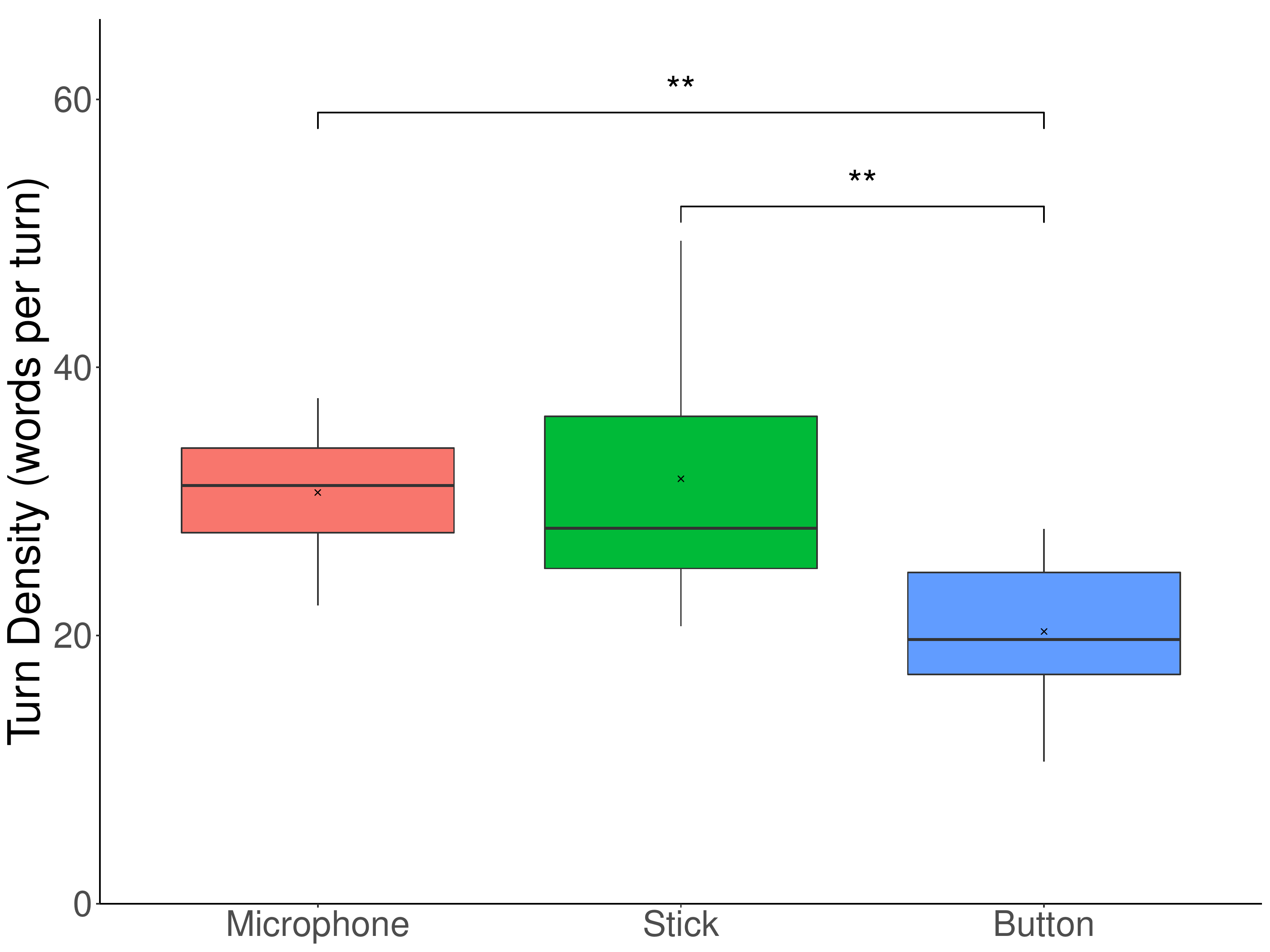}
    \caption{Plot of the conversation Turn Density (the mean number of words per turn).}
    \Description{Plot of the number of words per conversation turn.}
    \label{fig:wordsperturn}
\end{subfigure}
 \hfill
\begin{subfigure}{.47\textwidth}
  \centering
    \includegraphics[width=\textwidth]{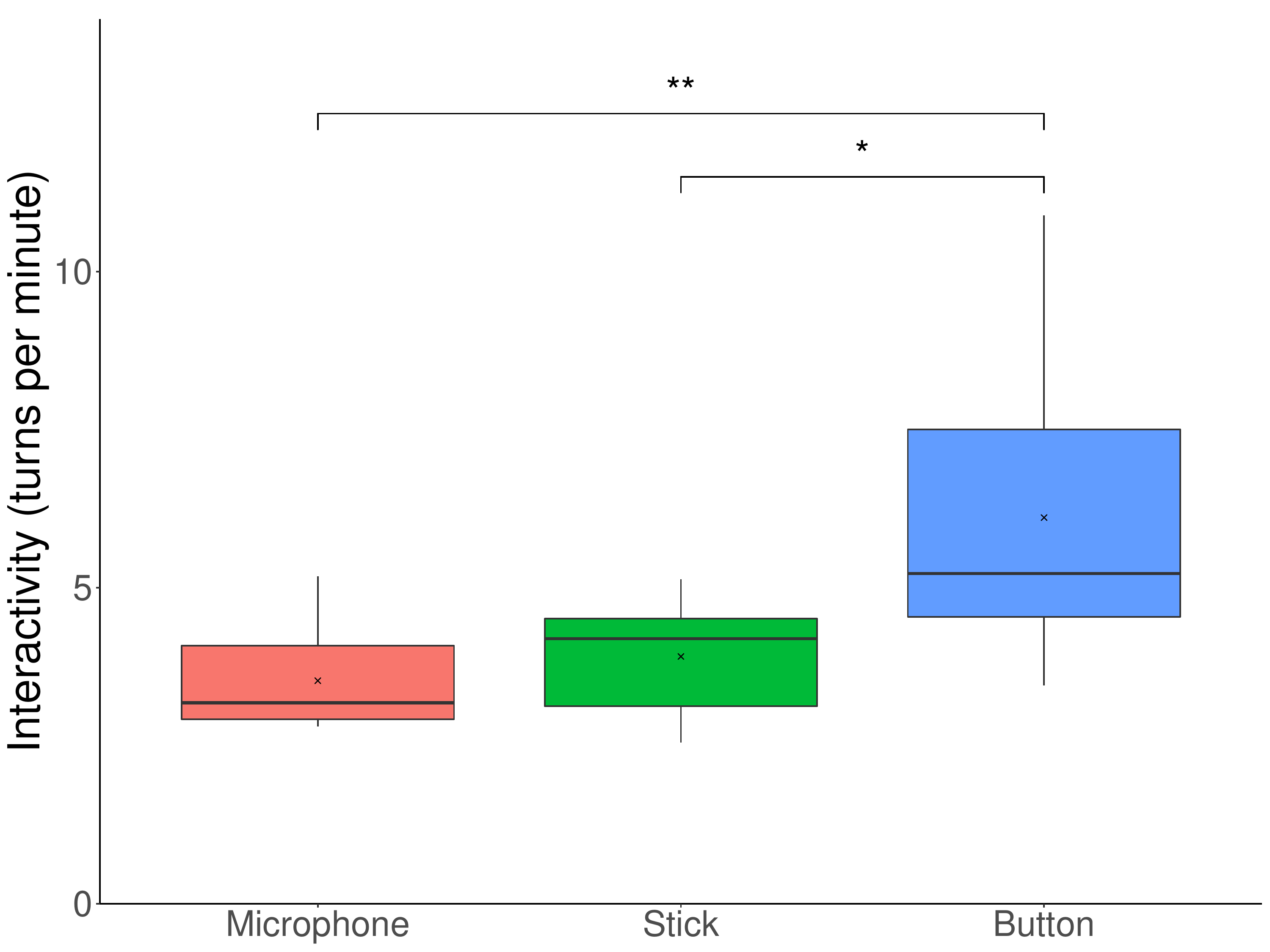}
    \caption{Plot of the conversation Interactivity (the number of conversation turns per minute).}
    \Description{Plot of the number of conversation turns per minute.}
    \label{fig:turnsperminute}
\end{subfigure}
 \hfill
\caption{Conversation Turn Density and Interactivity.}
\label{fig:convodynamics}
\end{figure}

We found a significant main effect ($F(2,27)=8.01$, $p<.01$) of \textsc{Condition} on conversation Turn Density (words per turn). We found that participants in the \textsc{Button} condition had significantly lower Turn Density than both the \textsc{Microphone} ( $p<.01$) and \textsc{Stick} ($p<.01$) conditions. For Interactivity (turns per minute), we found a significant impact of \textsc{Condition} ($F(2,27)=7.01$, $p<.01$). Participants in the \textsc{Button} condition had significantly higher Interactivity than both the \textsc{Microphone} ( $p<.01$) and \textsc{Stick} ($p<.01$) conditions. We found no significant effect of how long participants knew one another for either Turn Density ($F(3,26)=0.381$, $p=.768$) or Interactivity ($F(3,26)=1.02$, $p=.401$). The results, illustrated in \autoref{fig:convodynamics}, indicate that participants in the \textsc{Button} condition had shorter, more frequent speaker turns than in the other two conditions. The presence of the stick, on its own, did not significantly impact the conversation dynamics. For Speaking Ratio, we found no significant effect of \textsc{Condition} ($F(2,27)=0.117$, $p=.890$) nor how long participants knew one another ($F(3,26)=1.25$, $p=.312$).



\subsection{User Perceptions}
\begin{figure}[t]
    \centering
    \includegraphics[width=0.9\linewidth]{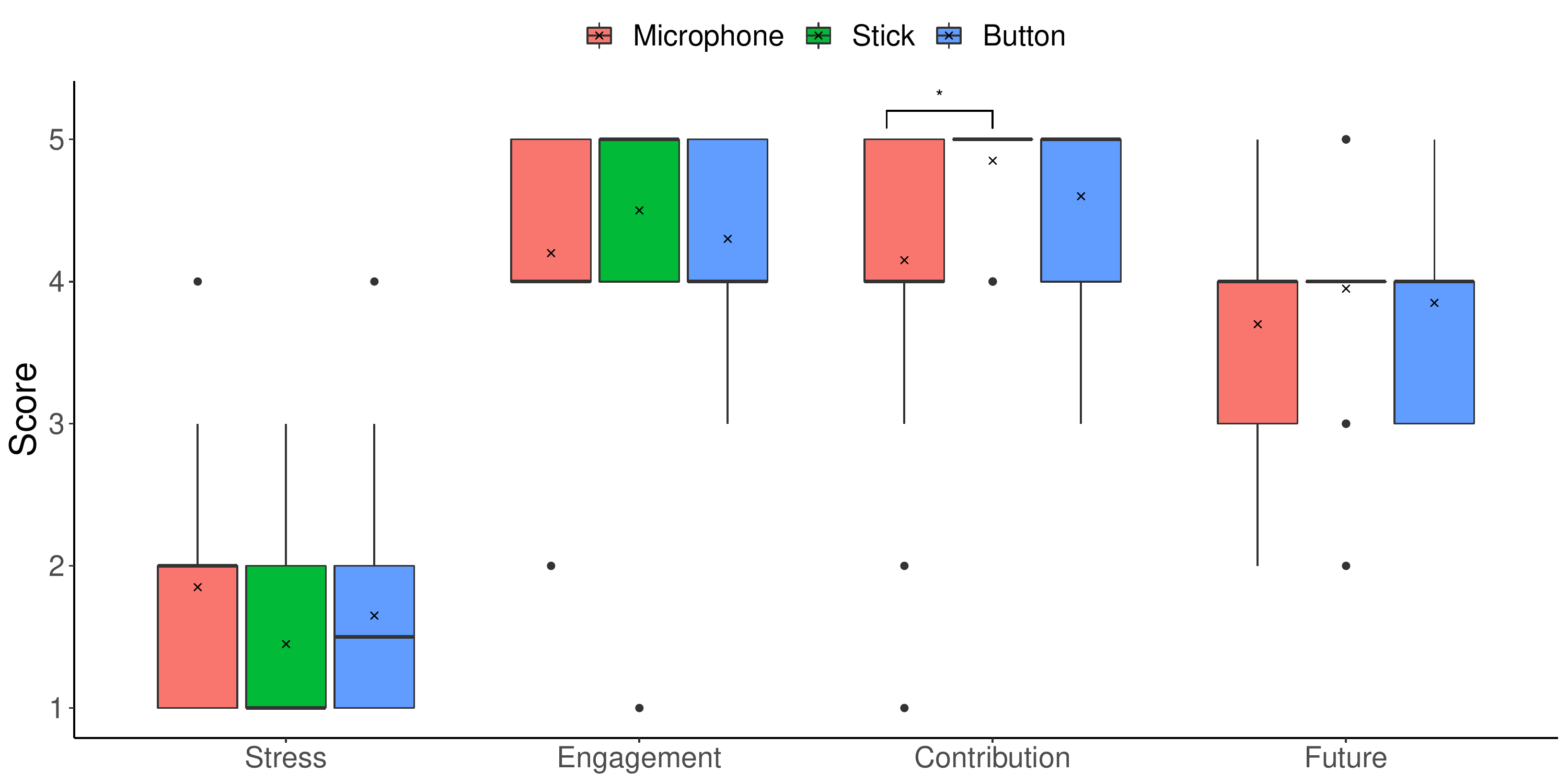}
    \caption{Plot of participants' perceptions of stress, engagement in conversation, contribution to the conversation, and the likelihood of using the system in the future for each condition. Possible scores range from 1 (low) to 5 (high) for each question. Responses to Contribution and Future have very low variance.}
    \Description{Plot of participants' perceptions of stress, engagement in conversation, contribution to the conversation, and the likelihood of using the system in the future for each condition. Possible scores range from 1 (low) to 5 (high) for each question. Responses to Contribution and Future have very low variance.}
    \label{fig:perceptions}
\end{figure}

We collected Likert-scale feedback on user perceptions of stress levels during the meeting (Stress), how engaged they were in the conversation (Engagement), how much they contributed to the conversation (Contribution), and the likelihood of using the system in the future (Future). The results are shown in \autoref{fig:perceptions}. Participants in all three conditions indicated low Stress and high ratings for Engagement, Contribution, and Future. We found a significant main effect of \textsc{Condition} on Contribution (ART-ANOVA results for all the perception questions are shown in \autoref{tab:anova_likert}). We found that participants in the \textsc{Stick} condition rated Contribution significantly higher than those in the \textsc{Microphone} condition ($p<.05$). 
Thus, participants perceived that they were contributing more to the conversation when using the prototype than users with unobtrusive microphones. We found no significant effect of how long participants knew one another for any of the questions ($p>.05$ for all).

\begin{table}[ht]
  \caption{ART-ANOVA results for all user perception Likert metrics}
  \label{tab:anova_likert}
  \begin{tabular}{cccc}
    \toprule
    \textbf{Question}   &\textbf{$df$}    &\textbf{$F$ value}   &\textbf{p}\\
    \midrule
    Stress              &(2, 57)        &$1.59$              &$.212$     \\
    Engagement          &(2, 57)        &$2.53$              &$.0883$   \\
    Contribution        &(2, 57)        &$4.27$              &$<.05*$    \\ 
    Use in Future       &(2, 57)        &$0.360$              &$.699$     \\
  \bottomrule
\end{tabular}
\end{table}


\subsection{Usability}

We administered the SUS to assess system usability in the three conditions. All three systems were rated as highly usable ($M_\textsc{Microphone}=81.5$, $M_\textsc{Stick}=83.75$, $M_\textsc{Button}=76.5$), and we found no significant effect of \textsc{Condition} on the SUS score using a one-way ART-ANOVA, $F(2,57)=1.46$, $p=.241$. We found no significant effect of how long participants knew one another ($p>.05$). These results suggest that differences in our results can be attributed to experiential differences between the systems rather than issues with any of the systems.

\begin{figure}[t]
\centering
    \centering
    \includegraphics[width=0.9\linewidth]{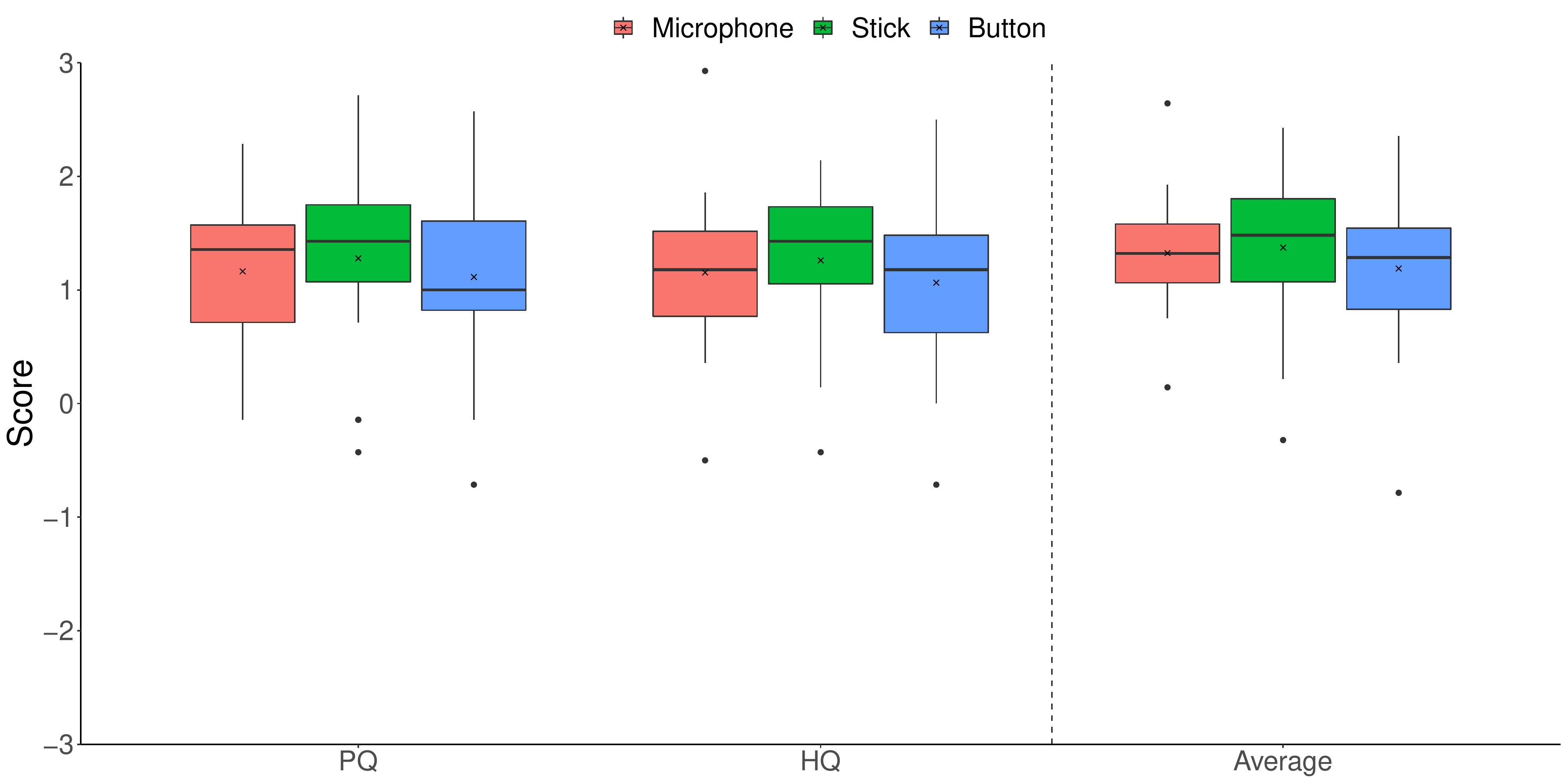}
    \caption{AttrakDiff results of the Pragmatic Quality (PQ) and Hedonic Quality (HQ), as well as the average.}
    \Description{AttrakDiff results of the Pragmatic Quality (PQ) and Hedonic Quality (HQ), as well as the average.}
\label{fig:attrakdiff}
\end{figure}

We used the AttrakDiff questionnaire to gain insight into the perceived user experience for each condition. The results are shown in \autoref{fig:attrakdiff}. We can see that all three systems were rated highly in each category, indicating that all three conditions elicited a positive user experience. There were no significant differences in the AttrakDiff metrics, as indicated by a one-way ART-ANOVA; pragmatic quality: $F(2,57)=0.385$, $p=.682$; hedonic quality: $F(2,57)=1.01$, $p=.372$; average: $F(2,57)=0.730$, $p=.487$. We found no significant effect of how long participants knew one another for any of the questions ($p>.05$ for all), suggesting that the user experience was not significantly impacted by inter-participant familiarity.

\subsection{Exit Interviews}
In the following, we present the results of the exit interviews clustered by themes identified during the coding process.

We first analyzed each interview to determine whether the participants felt they had adequately completed the assigned task (planning a workshop). All but one pair stated that they successfully completed the task and were satisfied with the outcome (100\% inter-rater agreement). The final pair had conflicting approaches to planning and had many disagreements, but they still conducted a complete meeting. Based on this, we conclude that our results should not be skewed by an inability to complete the assigned task.

\subsubsection{The Impact of Being Recorded}\label{sec:beingrecorded}
Although all participants were initially aware that they were being recorded, the recording technology often faded into the background: \textit{``it was non-existent. I didn't think one minute or one second that it was recorded''} (M04). This was mentioned by participants in all three conditions and was therefore true for both the clip-on microphones and the \prototype{} prototypes. The participants also noted that they did not feel that being recorded impacted their conversations: \textit{``I don't think it affected at all. It didn't disturb us in any way. I would say we didn't focus on the fact that we were being listened to''} (M09).

Furthermore, participants appreciated the relaxed atmosphere of the meeting, as they did not have to memorize what was said: \textit{``I feel like free to forget things because I didn't have this need of storing the main points of my conversation because that would have been big. I would have felt the need of writing it in my phone or something, but as I knew there was a device recording, I just ignored this need to remember things''} (S15).

There was a concern that a meeting transcript would require additional time after the meeting to process the information and create useful notes. Participants in the \textsc{Stick} condition reflected on post-processing the resulting transcript: \textit{``I would need to go through that conversation again, which is also time-consuming''} (S03). Participants in the \textsc{Button} condition were less concerned about post-processing and mentioned that the button created a more useful output: \textit{``it's actually not only just recording but also making a text out of it. That's pretty cool, but then also to highlight what I did. I think that at the end that might help to get back to the points which are important.''} (B04).

Finally, some participants also commented on the social acceptability of using a recording device in public. Participants felt that bystanders noticed them and their equipment during the meetings. This was true for groups carrying the stick, \textit{``Sometimes it felt maybe a bit weird to walk with that stick when other people were watching''} (S12), but also for groups with the clip-on microphones, \textit{``There were people who were looking at us, wondering what are we doing''} (M12).

Notably, there was no discernable pattern in sentiments towards being recorded based on how long participants knew one another. There was a mixture of lengths that pairs knew one another for both positive sentiments (e.g., forgetting being recorded) and negative sentiments (e.g., concern about bystanders).

\subsubsection{Stick Strategy}\label{sec:stickstrategy}
A single person carried the stick for nearly all pairs of participants in the \textsc{Stick} and \textsc{Button} conditions. Two pairs swapped halfway through the meeting, but no participants passed the stick back and forth more than once.

In line with participants forgetting that they were being recorded, several participants also mentioned that they forgot about the stick: \textit{``I forgot that I was holding it. I was just walking and it wasn't inconvenient''} (S05). However, participants were more likely to indicate that they were aware of the stick than the clip-on microphones. One effect of this awareness is that the presence of the system helped them focus on the meeting and stay on task: \textit{``When I was carrying it, it was a constant reminder that I'm on a serious work meeting situation and not just taking a walk with a friend''} (S04).

Participants were split on the stick form factor. On the one hand, some participants wanted something smaller and less obtrusive: \textit{``It would be nice if it were lighter [...] I thought it would be nice if you could stick it in your pocket and if it would be really light''} (S04). On the other hand, some participants acknowledged the benefits of a substantial artifact which requires intention: \textit{``not being so light makes it nicer, because it's not something you feel you can just throw, but it's something you have to take with you''} (S15).

\subsubsection{Button Strategy}\label{sec:buttonstrategy}
Almost all participants developed a similar strategy for using the button. For most participants, the person who carried the stick also pressed the button. However, for one pair, the person who was not carrying the stick controlled the button, and two pairs shared control of the button. There is no recognizable correlation between button strategies and how long the participants knew one another.

There were two main components to the strategies participants used to interact with the button. First, participants noted that they pressed the button when they made decisions or concluded on a point. Second, participants revealed that they would repeat or summarize important points before pressing the button so that the highlight in the transcript would capture the main idea: \textit{``when we were talking about a conclusion, then we pressed the button. We have a concrete conclusion on what we just discussed, and that was a summarization, and then the button is pushed, then we continue to the next topic''} (B03).

Some participants proposed the idea of individual buttons, which would enable users to have personalized notes: \textit{``additional buttons that the second or third person, or how many people there are, could run around with. I guess one stick or something and then multiple people that can press it. Then I take my notes and the other takes their notes''} (B07)


\subsubsection{Conversation Dynamics}
Most pairs of participants stated that they felt that the conversation was balanced, with both participants typically reporting that they had contributed somewhat equally to the meeting. However, there was a range of conversation dynamics present in the conversations. Many participants mentioned that one took the lead in organizing the conversation: \textit{``I was holding the paper [...], so I would say I was taking over the lead and asking the questions''} (M03). Many noted that their conversation flowed naturally and shifted back and forth: \textit{``it was mostly back and forth, but we were always on the same page, I would say. I think I started introducing a topic, and then we both liked the topic and then started planning. I would say we planned really naturally what would be the next step''} (S03). Some other participants assumed roles of idea generation and response: \textit{``I was giving him ideas, and he was trying to bring them up to life and to make them practical''} (M18). These different roles do not appear to be correlated with how long participants knew one another.

Participants had somewhat conflicting views on formality. Participants discussed many details in their planning meetings and felt that the recording technology would be useful for structured meetings: \textit{``It would be more for more structured meetings where you want to have fixed structured outcomes.''} (B01). However, participants still felt that a walking meeting would be inappropriate for more formal or important discussions: \textit{``I feel like if it's an important meeting, I wouldn't do it walking.''} (B18).

\subsubsection{Insights for Walking Meetings}
The participants consistently mentioned experiencing the expected benefits of walking meetings, indicating that the recording technology did not degrade the experience. Participants noted that they enjoyed being outside in the fresh air: \textit{``I liked that it was outside for sure. I liked that the weather was nice. I could get some fresh air, exercise a bit''} (M12). Participants also indicated feeling more concentrated, \textit{``I was concentrated on the content, actually only on the content because I was walking, so I had no time to think about anything else''} (M04), and more creative, \textit{`` I think it will always be easier to think about creative ideas in coming up with things when you are walking''} (S13). Participants also noted a feeling of efficiency because they were accomplishing two things at once by being productive while doing physical activity: \textit{``My brain was still working that I could be productive at that time [...] was nice to have a walk and still be productive. I think it like that. Using time efficiently.''} (S03) 

Some participants found some aspects of the environment distracting, \textit{``Since we are in the park, there are some people, I could see, for example, a dog, which was taking a bath in the pond, and this distracted me''} (M12). Others, however, mentioned that they were less distracted outside than they would be in their office: \textit{``I was entirely focused on the meeting and on the conversation. I was walking almost automatically, and there were no distractions, actually. I think that there were less distractions than while I'm sitting in the office''} (M15).

Participants also commented on several ways in which the physical space played a role in their meetings. The participants used physical cues to indicate when the meeting should end: \textit{``it was [a] very concrete way to end the meeting that we're coming to the end of the walk, the meeting is done''} (B14). Beyond such practical applications, the changing environment also helped with idea generation and creativity: \textit{``it opens the mind because the environment changes and we have the possibility to be more creative''} (M11). Finally, conducting the meeting in a changing environment allowed participants to use spatial cues to remember points from the meeting: \textit{``I remember \textbf{where} we talked about it, and I can easily remember what we actually talked about without actually having access or needing access to the transcription''} (B05).

Some participants used physical movement as an additional modality for expression in their meetings. For example, participants moderated their speed based on their emotions: \textit{``depending on how we talk, we could also moderate our walk. We could also react to this. A static position would not help us here [...] When we have more energy, like for example, we're angry or happy about something, we tend to give away more energy. It was also nice to see how I reacted and when I wanted to walk faster and when the other person also reacted when he wanted to go faster, move on, or when he slowed down. It is just something that we could also moderate''} (M07).


\section{Discussion}
We conducted this investigation to explore the research question: \textit{How does a shared tangible recording artifact impact walking meetings?} In the following, we will discuss insights gained from the \prototype{} for transcription-based note-taking technologies and meetings in motion.

\subsection{Interpreting the Results}
All three systems were rated as highly usable (SUS) and elicited a positive user experience (AttrakDiff). In line with these results, participants in all conditions had low stress levels during the meeting and rated their engagement in the conversation and contribution to the meeting as high. We also found that participants were highly successful in completing the assigned task and were highly likely to use the system in the future. Participants consistently experienced many of the established benefits of walking meetings (e.g., increased creativity), suggesting that our support systems do not degrade the walking meeting experience. We also found that the highlighting button had additional impacts on conversation metrics, leading to shorter and more frequent conversations and more interaction. Notably, the length of time participants knew one another had no impact on any other variables, although this is mitigated by the fact that all participants did know one another before the study. Overall, these results indicate that transcription-based note-taking systems, in general, can be effectively used during walking meetings and generate interesting new conversation dynamics. Based on this finding, future work should investigate whether such systems incentivize walking meetings over the long term.

\subsection{Recording Gives Confidence for More Complex Meetings}
One of the goals of this research is to investigate ways to make walking meetings more attractive and applicable to a wide array of meetings. Prior work has consistently found that people associate walking meetings with a limited subset of meetings, typically brainstorming and informal discussions~\cite{haliburton_charting_2021, damen_understanding_2020}. While some prior work uses the persuasive approach (e.g., ~\cite{ahtinen_brainwolk_2016, ahtinen_lets_2017, ahtinen_walk_2016}) to nudge people into conducting more walking meetings, we take an approach more similar to \citet{damen_hub_2020} and \citet{haliburton_charting_2021}, who suggest using technology to make walking meetings more convenient and practical. In this study, our users remarked that using recording technology made them feel free to forget (see Section \ref{sec:beingrecorded} (S15) since they were not worried about their inability to take notes. Additionally, participants in the \textsc{Button} condition noted that the highlights resulted in more useful notes than a full transcript. Our results suggest that users should be able to conduct more complex and detailed meetings while walking if they use recording and highlighting technology similar to the \prototype{}. We investigated our system with planning meetings, where users discussed fine details of a future event, which already moves beyond the traditional domain of walking meetings (i.e., early brainstorming and informal discussions). While we do not suggest that \textit{every} meeting is suitable to be conducted in motion, further research is required to quantify the degree to which such technologies can open up the possibility space for walking meetings. \citet{allen_understanding_2014} developed a 16-category taxonomy of meeting purposes, which could provide a useful framework for future investigations into the applicability of walking meetings for different meeting types.

Our study also highlights an interesting contradiction. Past work shows that one of the main barriers for walking meetings is the lack of ability to take notes~\cite{haliburton_charting_2021, damen_understanding_2020}. In our study, participants had notes they found useful and generally described their meetings as very successful. Despite these successes, participants still emphasized that they felt that walking meetings are only appropriate for informal brainstorming meetings. Shifting this conventional perspective on walking meeting applicability likely requires repeatedly using the technology to establish trust in the effectiveness and usefulness of the resulting transcripts. Participants also mentioned some additional desires for the technology, such as support for displaying visuals or automatic summaries of the transcripts, which could further increase the utility of the prototype and increase the complexity of meetings that would be feasible to conduct while walking.

Recording walking meetings also raises certain ethical considerations. Since walking meetings generally occur in public spaces, the recording device may accidentally capture bystander conversations. The ethics of recording in public and an individual's right to privacy has been discussed in prior work on lifelogging~\cite{wolf_lifelogging_2014}. This is an important consideration because bystanders have not consented to be recorded. One practical way to combat this would be for the recording system to automatically remove content generated by any speaker other than the main participants in the meeting. This scheme could ensure that bystander conversations never appear in the transcript.

\subsection{A Physical Artifact Communicates Shared Understanding}
Participants in the \textsc{Microphone} condition were more likely to completely forget about the system, while participants with the stick noted that the physical artifact helped people focus. In particular, \textsc{Stick} and \textsc{Button} participants identified that the artifact helped them stay on task and signified that they were conducting a serious meeting. Prior work in HCI, such as MemStone~\cite{bexheti_memstone_2018}, has shown that physical artifacts create a shared understanding that a meeting has begun and can increase concentration. Interestingly, participants in the MemStone study noted that being recorded would change their behavior, while our participants consistently stated that recording had little to no impact on their conversations aside from staying on task. As the largest difference between the two studies is the fact that our meetings were conducted while walking, it may be that the additional stimulation and changing environment of walking meetings influenced this perception. 

Despite recognizing the benefits of the tangible device, participants were mixed on whether they wanted a smaller, less obtrusive device. This may be an example of convenience trumping effectiveness~\cite{dhillon_when_2012} but is also likely influenced by the fact that the device is novel and ubiquitous capture devices are not yet commonplace in society. One potential alternative would be to record directly with a smartphone, as they are already ubiquitous. However, the smartphone would need to be held by a participant to capture the audio from both users, which may cause users to be distracted by notifications or the phone itself since the mere presence of a smartphone is distracting~\cite{thornton_mere_2014,allred_mere_2017}. Smartphone microphone technology may progress in the future to the point where a conversation could be recorded from a user's pocket, which would then be similar to our unobtrusive \textsc{Microphone} condition but with a more convenient initial setup. However, some participants also mentioned that they appreciated the large object because the added tangibility adds intention. This mixed opinion on the device size may also be an indication that different physical artifact forms may be appropriate for different users and perhaps also for different meetings. Some users mentioned that people were looking at them when they carried the prototype, so a small unobtrusive device may be more appropriate for locations with many bystanders (e.g., city sidewalks), while a larger device would be acceptable in less crowded locations (e.g., in a park). Past work has identified social acceptance as a barrier to walking meetings since they can be perceived by others as simply going outside for a break~\cite{damen_understanding_2020}. One motivation for a larger physical form would be for workers to signal to colleagues that they are going outside for a serious work-related task.

\subsection{A Highlighting Button Facilitates Turn-Taking and Summarization}
We found that introducing a highlighting button significantly decreases turn density and increases interactivity during walking meetings. This means that participants in the \textsc{Button} condition spoke in shorter bursts and exchanged speaking roles more often during their meetings. Past research on turn-taking in the medical field by \citet{roter_interactive_2008} recommends increasing interactivity as a method of improving physician-patient conversations, as having both parties participate more in the conversation leads to increased understanding. Viewed through this lens, our results indicate that a highlighting button is an effective way to improve conversation dynamics and democratize meetings.

Although we cannot definitively say why the highlighting button had this impact, there are several factors that may have influenced participants' conversation dynamics. The button is a tangible object which represents ideas being logged into the meeting record. As such, participants may have felt that they were sharing in the creation of a record of their thoughts, leading to a more rapid exchange of ideas. Nearly all participants in the \textsc{Button} condition indicated that their strategy for using the button was to press it when they concluded a decision or discussion point. This action potentially facilitated progress through the meeting agenda and adherence to structure, as pressing the button was a concrete and tangible indication that a point was finished and participants could move on with the discussion. Our study was not designed to measure task progress or accomplishment, so we have no indication of whether participants in the \textsc{Button} condition actually accomplished more or simply exchanged speaking roles more often. A future study where the meeting task is designed to measure whether a highlighting button enables participants to accomplish more during a meeting would be highly informative.

Past work in HCI has also investigated turn-taking and coordination in remote meetings~\cite{bohus_multiparty_2011}. \citet{chan_designing_2008} found that multi-modal (i.e., haptic and visual) icons were the preferred manner of facilitating turn-taking. Future work should investigate whether our findings could be extended by integrating similar haptic icons to enhance turn-taking coordination during walking meetings.

The button also impacted the way our participants conducted their conversations. Nearly all participants developed a similar strategy, whereby they repeated statements and summarized their points before they pressed the button. This behavior mimics typical text-based note-taking techniques and recommended note-taking methods, such as outlining and summarizing key points~\cite{kiewra_providing_1988}. Research by \citet{choi_mnemonic_2017} shows that summarizing and re-stating are mnemonic strategies that reinforce collective memory and information distribution. The techniques adopted by participants in the \textsc{Button} condition are therefore aligned with recommended strategies to improve the effectiveness and impact of meetings. Past work in CSCW by \citet{kalnikaite_markup_2012} used a similar button technique and found that it decreased load, increased conversation contributions, and improved recall two weeks after the meetings.

\subsection{The Talking Stick Metaphor}
Although the button promoted turn-taking, no participants adopted the metaphor whereby only the holder of the talking stick is allowed to speak, which is a primary function of some traditional talking sticks (e.g., \cite{wolf2003talking}). As noted in Section \ref{sec:task}, the participants were free to choose who carried the stick and whether they passed it back and forth. We chose not to enforce the traditional talking stick paradigm because we wanted to investigate how the prototype naturally impacted the conversation dynamics. The traditional talking stick rules may not be appropriate for every type of meeting and was not the subject of this investigation. Although this result was not surprising, it is notable that the vast majority of participants did not distribute the stick carrying role at all. Only two pairs swapped the stick partway through (as reported in Section \ref{sec:stickstrategy}). This result can be partially explained by the fact that many participants forgot they were holding the stick, and most participants in the \textsc{Stick} and \textsc{Button} conditions mentioned that they were very focused on the meeting. While our design was inspired by the concept of a talking stick, it was not our aim or expectation for the device to induce talking stick rules.

Beyond denoting whose turn it is to speak, talking sticks are also used to indicate that an important conversation is taking place. This theme, which is more aligned with our design intention, was acknowledged by the participants. As previously discussed, the presence of the stick communicated a shared understanding between participants that a serious meeting was taking place and attention should be focused on the task at hand. This reveals that although the participants did not literally share the object, in that they did not share the responsibility of carrying the object, the physical artifact was still shared through signaling and mutual understanding. Past work has shown that tangibles can signal various meanings in social communication~\cite{shaer_tangible_2010}, from representing responsibility or supporting ambient awareness.

\subsection{Ways Forward}\label{sec:waysforward}
We highlight potential ways forward for designing technologies to support note-taking while in motion based on our results. The tangible, physical form of the \prototype{} had a positive impact by conveying a shared understanding, helping users stay on task, and communicating that they were participating in a serious meeting. However, the prototype could be further developed to exploit the benefits of tangible interaction. Future work should investigate additional affordances available with the walking stick format. For example, users could tap the stick on the ground, spin it, flip it upside down, or pass it back and forth as input modalities. Vibration motors or lights could be added for additional feedback to the user. Implicit interaction could also be incorporated, such as highlighting text based on changes in walking speed. Additional research is required to understand how such gestures, feedback modalities, and implicit interactions might impact the meeting experience.

Several participants mentioned a desire for each meeting member to have their own highlighting button (Section \ref{sec:buttonstrategy}). This could be implemented in several ways to generate more useful notes. In the most basic format, all users would contribute to a single set of shared highlights. The input could also be labeled, so participants could see which parts of the meeting were marked as important by which participants. Segments that are highlighted by more than one person could be emphasized as particularly important. In a different implementation, highlights could be private. Each person would only see the sections that they highlight. A hybrid version could also be created where highlights are private by default, but if multiple people highlight the same section, this information would be communicated to all. It is likely that the different modes would be appropriate for different meetings and different participants, so these modes should be personalizable.

On the subject of privacy, although it was not mentioned by our participants, we foresee that there may be situations where users want to discuss something `off the record' as discussed by Haliburton et al.~\cite{haliburton_charting_2021}. It would therefore be prudent for future iterations to include an easy method of pausing the recording function or wiping a previous statement from the recording. Given that participants summarized and restated points when they used the highlighting button, a privacy-forward approach would be for the system to only record while the button is pressed. Alternatively, the system could always be recording and storing a short buffer which automatically deletes itself if the button is not pressed. If the button is pressed at some point, the buffer is saved. This configuration would enable users to retroactively capture important points after they are said, while otherwise deleting content and maintaining privacy. The challenge with either privacy-forward design is that the users need to remember to press the button, or else nothing is recorded. In the current design, even if the users forget to take highlights, they still receive a full transcript which they can annotate after the fact. Future research is required to investigate whether the benefits of the privacy-forward approach outweigh the potential usability issues.

The \prototype{} concept could also be applied to technology-supported remote walking meetings. \citet{haliburton_charting_2021} developed design fictions where users in disparate geographical locations could meet while walking. Each participant could bring a \prototype{} with them to record the meeting and generate highlights. Since the user needs in this situation are different than in a co-located meeting, a different physical artifact may be appropriate. Future work is required to generalize our concept to remote meetings, which would greatly expand the applicability of walking meetings.

\subsection{Limitations}
One limitation of our work is the fact that users were planning a fictitious event. We conducted an outdoor user study where every pair of participants was given the same prompt to conduct their meeting, which was chosen to reduce variability in meeting content. However, it would be highly informative for future studies to investigate similar systems in the wild. An in-the-wild study with real meetings would likely have higher stakes, and therefore the transcripts generated by the system would be more valuable and, therefore, more closely scrutinized by participants.

As with any study involving human-human interaction, there are many uncontrollable factors. We recorded how long each pair of participants knew one another and found no significant correlation with any of our metrics. However, there are other factors that could have impacted the interaction, such as how extroverted, talkative, or domineering each participant is. Compared to average sample sizes at CHI~\cite{Caine2016}, we collected a relatively large sample to combat this variability and found no significant difference in Speaking Ratio between conditions, so we assume that the distribution of such personality traits is random.

Another potential limitation is that we did not investigate a scenario with unobtrusive microphones and a highlighting button. Although this, theoretically, would have resulted in a two-by-two study design (tangible object vs. unobtrusive and highlighting vs. no highlighting), this would not work in practice. Introducing a highlighting button to the unobtrusive microphone condition would inherently introduce a new physical object. On the one hand, this would, therefore, no longer be a non-tangible condition. On the other hand, this would require an entirely new design process. For example, should each participant have a small button or should there be some shared object? Should the button be a physical object that it only used for highlighting, or should participants directly input highlights on a mobile phone? As discussed in Section \ref{sec:waysforward}, it would be interesting for future work to investigate individual highlighting buttons and shared control paradigms.

Finally, many of our participants do not regularly conduct walking meetings, and it is currently not common practice to record audio transcripts of meetings. Therefore, there was likely a double novelty effect resulting from the combination of meeting while walking outside and using audio transcription. It would be useful for future research in this field to investigate how user perceptions change and stabilize when using such a system over the long term.

\section{Conclusion}
In this paper, we investigated a shared tangible artifact and a shared highlighting button for automatically capturing notes during walking meetings. We designed two proof-of-concept versions of the \prototype{} and evaluated them in a between-subjects study with 60 people comparing three conditions: \textsc{Microphone} (baseline with unobtrusive clip-on microphones), \textsc{Stick} (the \prototype{} without a highlighting button), and \textsc{Button} (the \prototype{} with a highlighting button). Our results show that the \prototype{} increased task focus and created a shared understanding between users. The  addition of the shared highlighting button fostered new conversation dynamics and mnemonic strategies and generated more useful notes. We contribute insights for future systems to create notes on the move using recording and transcription technologies. By further developing this field, we hope to expand the scope of walking meetings to include more complex, note-heavy meetings. Through technology-supported walking meetings, we aim to provide an opportunity for users to integrate physical activity into productive workday routines, thereby improving the overall health of the workforce.


\begin{acks}
This work was supported by the Bavarian Research Alliance association ForDigitHealth, by the European Union’s Horizon 2020 Programme under ERCEA grant no. 683008 AMPLIFY, and by the Swedish Research Council award number 2022-03196.
\end{acks}

\bibliographystyle{ACM-Reference-Format}
\bibliography{references}

\end{document}